\documentclass[aip,twocolumn,reprint,showpacs,superscriptaddress]{revtex4-1}
\usepackage{amsmath}
\usepackage[latin9]{inputenc}
\usepackage{graphicx}
\usepackage{hyperref}
\usepackage{siunitx}
\usepackage{graphicx}
\usepackage[color= blue!20]{todonotes}
\usepackage[normalem]{ulem}
\usepackage{color}

\newcommand{\ket}[1]{\left| #1 \right\rangle} 

\begin{document}

\title{Coupling ultracold atoms to a superconducting coplanar waveguide resonator}

\author{H. Hattermann}\email{hattermann@pit.physik.uni-tuebingen.de}\affiliation{CQ Center for Quantum Science in LISA$^+$, Physikalisches Institut, Eberhard Karls Universit\"at
T\"ubingen, Auf der Morgenstelle 14, D-72076 T\"ubingen, Germany}
\author{D. Bothner}
\affiliation{CQ Center for Quantum Science in LISA$^+$, Physikalisches Institut, Eberhard Karls Universit\"at
T\"ubingen, Auf der Morgenstelle 14, D-72076 T\"ubingen, Germany}
\affiliation{Present Address: Kavli Institute of Nanoscience, Delft University of Technology, PO Box 5046, 2600 GA Delft, The Netherlands}
\author{L. Y. Ley}\affiliation{CQ Center for Quantum Science in LISA$^+$, Physikalisches Institut, Eberhard Karls Universit\"at
T\"ubingen, Auf der Morgenstelle 14, D-72076 T\"ubingen, Germany}
\author{B. Ferdinand}\affiliation{CQ Center for Quantum Science in LISA$^+$, Physikalisches Institut, Eberhard Karls Universit\"at
T\"ubingen, Auf der Morgenstelle 14, D-72076 T\"ubingen, Germany}
\author{D. Wiedmaier}\affiliation{CQ Center for Quantum Science in LISA$^+$, Physikalisches Institut, Eberhard Karls Universit\"at
T\"ubingen, Auf der Morgenstelle 14, D-72076 T\"ubingen, Germany}
\author{L. S\'ark\'any}\affiliation{CQ Center for Quantum Science in LISA$^+$, Physikalisches Institut, Eberhard Karls Universit\"at
T\"ubingen, Auf der Morgenstelle 14, D-72076 T\"ubingen, Germany}
\author{R. Kleiner}\affiliation{CQ Center for Quantum Science in LISA$^+$, Physikalisches Institut, Eberhard Karls Universit\"at
T\"ubingen, Auf der Morgenstelle 14, D-72076 T\"ubingen, Germany}
\author{D. Koelle}\affiliation{CQ Center for Quantum Science in LISA$^+$, Physikalisches Institut, Eberhard Karls Universit\"at
T\"ubingen, Auf der Morgenstelle 14, D-72076 T\"ubingen, Germany}
\author{J. Fort\'agh}
\affiliation{CQ Center for Quantum Science in LISA$^+$, Physikalisches Institut, Eberhard Karls Universit\"at T\"ubingen, Auf der Morgenstelle 14, D-72076 T\"ubingen, Germany}
\begin{abstract}

We demonstrate coupling of magnetically trapped ultracold $^{87}$Rb ground state atoms to a coherently driven superconducting coplanar resonator on an integrated atom chip.
We measure the microwave field strength in the cavity through observation of the AC shift of the hyperfine transition frequency when the cavity is driven off-resonance from the atomic transition.
The measured shifts are used to reconstruct the field in the resonator, in close agreement with transmission measurements of the cavity, giving proof of the coupling between atoms and resonator.
When driving the cavity in resonance with the atoms, we observe Rabi oscillations between atomic hyperfine states, demonstrating coherent control of the atomic states through the cavity field.
The observation of two-photon Rabi oscillations using an additional external radio frequency enables the preparation of magnetically trapped coherent superposition states near the superconducting cavity, which are required for the implementation of an atomic quantum memory.

\end{abstract}
\pacs{}
\maketitle
\section*{Introduction}
\vspace{-2mm}
Hybrid quantum systems of superconductors and atomic spin ensembles have been proposed \cite{Xiang2013, Andre2006, Henschel2010} for quantum information processing to overcome the limited coherence of superconducting qubits \cite{Kim2011, Paik2011}.
In the envisioned hybrid system, information is processed by fast superconducting circuits and stored in a cloud of cold atoms, which serves as a quantum memory \cite{Verdu2009,Patton2013, Patton2013a}.
Information is transferred between the two quantum systems using a superconducting coplanar waveguide resonator as a quantum bus.
In recent years, coupling between superconducting structures and spin-systems such as nitrogen vacancy centres \cite{Kubo2010, Kubo2011, Amsuess2011, Putz2014, Grezes2016} and ions in solid state systems \cite{Schuster2010, Probst2013} has been observed.
Cold atoms coupled to superconducting resonators would furthermore enable the implementation of novel quantum gates \cite{Petrosyan2009, Petrosyan2008, Pritchard2014, Sarkany2015}, the realization of a microwave-to-optical transducer\cite{Hafezi2012} and on-chip micromasers \cite{Yu2017mm}.
The interaction between Rydberg atoms and three-dimensional superconducting microwave resonators has been a rich research topic, especially with regard to atom-photon interactions on the fundamental level \cite{Haroche2006}.
Research on planar superconducting structures, however, holds the promise of switchable interactions between the subsystems, integration with scalable solid-state circuitry \cite{Wallraff2004, DiCarlo2010, Lucero2012} and long information storage in the atomic ensemble.  
While long coherence times in cold atoms have been studied extensively \cite{Treutlein2004, Deutsch2010, Kleine2011, Dudin2013, Bernon2013} and trapping and manipulation of atoms in the vicinity of superconducting chips has been demonstrated in a series of experiments \cite{Nirrengarten07, Mukai07, Roux08, Minniberger2014, Mueller2010, Weiss2015}, coupling between trapped atoms and planar superconducting resonators has not been shown yet. 

In this article, we demonstrate magnetic coupling of ultracold magnetically trapped atoms to a superconducting coplanar waveguide resonator operated at temperatures around 6\,K. 
The cavity is near-resonant with the atomic hyperfine splitting of $^{87}$Rb and coherently driven by an external microwave synthesizer.
We investigate both the dispersive as well as the resonant coupling regime. 
By driving the cavity off-resonantly with respect to the atoms, the atomic states reveal an AC-Zeeman shift under the influence of the microwave (MW) field \cite{Sarkany2014}.
This leads to a shift of the atomic transition frequency, which is measured by Ramsey interferometry. 
We use the AC-Zeeman shift to reconstruct the microwave intensity in the coplanar resonator. 
In contrast, when the cavity is driven at a frequency corresponding to an atomic transition, Rabi oscillations between atomic hyperfine states are observed.

Our measurements present a vital step towards the realization of a atom-superconductor hybrid system, paving the way towards the implementation of an atomic quantum memory coupled to a superconducting quantum circuit and the realization of microwave-to-optical transducers.

\section*{Results}
\vspace{-2mm}
\subsection*{Atomic ensembles trapped in a coplanar waveguide resonator}
\vspace{-2mm}
\begin{figure*}
\centerline{\includegraphics[width=1\textwidth]{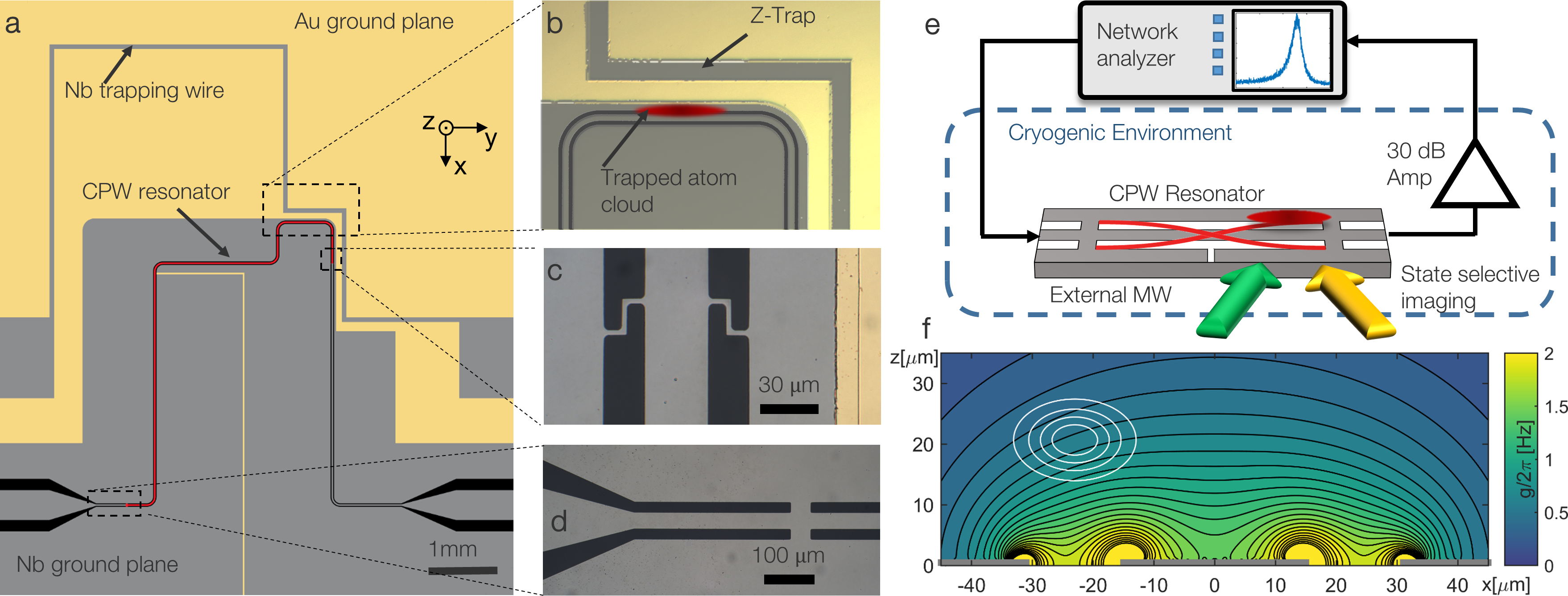}}
\caption{\textsf{\textbf{A superconducting atom chip for coupling ultracold atoms to a coplanar resonator. a} Schematic top view of the superconducting atom chip, comprising a Z-shaped trapping wire and a coplanar microwave resonator (centre conductor marked in red). Parts of the niobium ground planes have been replaced by gold to circumvent the Meissner effect and facilitate magnetic trapping. The slit in the lower ground plane prevents the formation of a closed superconducting loop. \textbf{b} Optical microscope image of the trapping region with the position of the atoms trapped close to the antinode of the resonator. During the measurements, trapping is purely provided by persistent supercurrents around the upper cavity gap and external fields. \textbf{c} Microscope image of the coupling inductances at the output of the resonator and \textbf{d} at the input of the resonator. \textbf{e} Scheme of the measurement setup. Atoms are coupled to a driven coplanar waveguide resonator and detected by state-selective absorption imaging. For Ramsey experiments in the dispersive regime, additional external microwave fields are used to manipulate the atoms. \textbf{f} Simulated coupling strength between a single ground state atom and a single photon in the cavity, resonant to the $\left|1, -1 \right\rangle \rightarrow \left|2,0 \right\rangle$ transition. The white lines indicate positions of equal density for an atomic cloud of temperature $T_\text{at} = 800$\,nK in the trap, corresponding to 20, 40, 60, and 80\% of the density in the centre.}}
\label{fig:chip}
\end{figure*}
For our experiments, we magnetically trap an ensemble of ultracold $^{87}$Rb atoms in the state  5S$_{1/2} F= 1, m_F = -1:=\ket{1,-1}$ close to a coplanar microwave resonator on an integrated atom chip.
The chip comprises two essential structures: a Z-shaped wire for magnetic trapping of neutral atoms and a superconducting coplanar waveguide (CPW) resonator (Fig.\,\ref{fig:chip}).
The CPW resonator is an inductively coupled half wavelength cavity \cite{Bothner2017} with a fundamental mode resonance frequency $\omega_\mathrm{Res} \approx 2\pi\cdot6.84\,$GHz and a linewidth of $\kappa \approx2\pi\cdot 3\,$MHz in the temperature range ($T = 6-7\,$K) relevant for the experiments described here.
By varying the temperature of the atom chip, the resonance frequency of the microwave cavity can be tuned by about $30\,$MHz, where the atomic hyperfine transition frequency $\omega_\mathrm{HF} = 2\pi \cdot 6.8347\,$GHz lies within this tuning range.
Details on the chip design, fabrication methods and cavity properties can be found in sections S1-S3 of the Supplementary Information to this article. 
With the coupling inductors (Fig.\ref{fig:chip}c and d) the microwave cavity gap close to the Z-trap provides a closed superconducting loop on the chip, in which the total magnetic flux is conserved.
The other resonator gap does not form a closed loop, as the lower ground plane has been cut to avoid flux trapping.
We take advantage of the flux conservation by freezing a well-defined amount of magnetic flux into the closed loop during the chip cool-down.
A conservative magnetic trapping potential for the Rb atoms in the vicinity of the cavity mode is formed by the combination of flux conserving loop currents and a homogeneous external field  \cite{Bothner2013, Bernon2013}.
A homogeneous offset field along the $y$-axis $B_\mathrm{off} = 0.323\,$mT is additionally applied to ensure a non-zero magnetic field amplitude in the trap minimum to avoid spin-flip losses.
$N_\text{at}\sim 10^5$ atoms are magnetically trapped at a distance of $\sim 20\,\mu$m above one of the coplanar waveguide gaps and close to one of the ends of the cavity, where the antinodes of the standing microwave magnetic fields are located, cf.\,Fig.\,\ref{fig:chip}b and e.
At this position, the magnetic microwave field of the transversal wave in the cavity is oriented perpendicular to the quantization axis of the atomic spins ($y$-direction).
Figure \ref{fig:chip}f depicts the coupling to the magnetic microwave field of the cavity, obtained from finite element simulations (see Supplementary section S4), in a cross-sectional view of the resonator.
Solid white lines indicate the calculated positions of equal atomic density for an atomic cloud of 800\,nK. 
From the microwave field amplitude at the position of the atoms we estimate an average single-atom single-photon coupling strength of $g = \vec{\mu}\cdot\vec{B}_\text{ph}\approx 2\pi\cdot0.5\,$Hz. 
The magnetic microwave field and thus the coupling can be considered constant along the atomic cloud with an extension of $\sim100\,\mu$m in $y$-direction, which is about two orders of magnitude smaller than the cavity and thus the wavelength.
For the experiments described in this article, the cavity is driven by an external microwave synthesizer. 
In the limit of high photon numbers $n_\text{ph}\gg N_\text{at}$ explored in this article, the cavity field can be treated classically, and the collective coupling between an atom and the cavity is small compared to the damping rate.
In the classical regime, the atoms couple individually to the cavity field, hence the Rabi frequency is independent of the number of atoms in the cavity \cite{Chiorescu2010}.

\subsection*{Sensing the cavity field with cold atoms}
\vspace{-2mm}
\begin{figure*}
\centerline{\includegraphics[width=\textwidth]{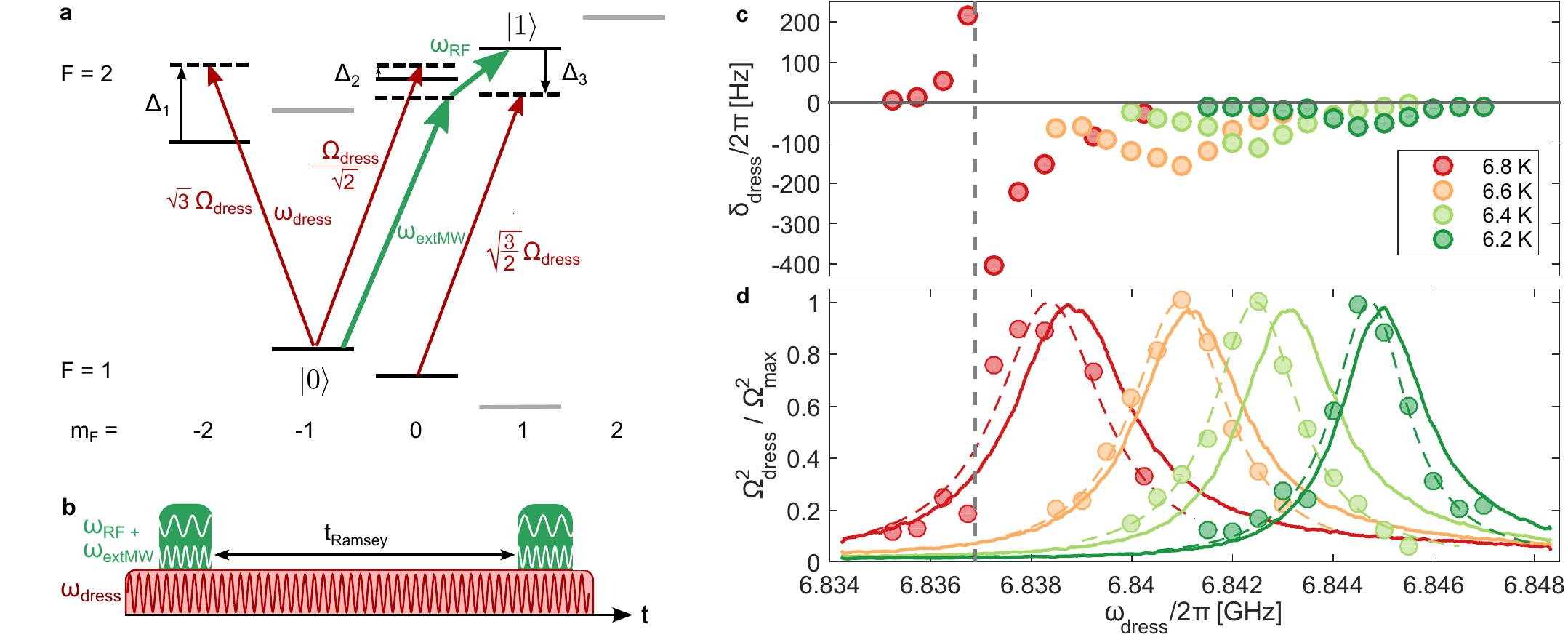}}
\caption{\textsf{\textbf{Probing the off-resonant cavity field with cold atoms. a} Level scheme of the $^{87}$Rb ground state manifold. Shown are the external MW and RF frequencies used for driving the two-photon transitions for the Ramsey scheme (green) and the off-resonant coupling of the cavity field to the relevant states (red). \textbf{b} Experimental timing for the Ramsey sequence. The cavity field (red) is driven throughout the interferometric sequence. \textbf{c} Measured shift of the Ramsey frequency vs frequency of the field in the superconducting microwave resonator for different chip temperatures.
The sign change in the 6.8\,K curve occurs at crossing the $\ket{1,0} \rightarrow \ket{2,1}$ transition, i.e. when $\Delta_3 = 0$, as indicated by dashed vertical line. \textbf{d} Data points: Calculated microwave intensity $\Omega^2_\text{dress}$ based on the measurements of $\delta_\text{dress}$. The coloured dashed lines are Lorentzian fits to the data points. The solid lines are the measured transmission spectra of the microwave resonator.}}
\label{fig:Ramsey}
\end{figure*}
When driving the resonator at a frequency $\omega_\text{dress}$ off-resonant to the atoms, the atomic transition is shifted by the MW field.
This AC-Zeeman shift can be experimentally detected and used to reconstruct the intensity of the cavity field.
We measure the frequency of the atomic transition between the magnetically trapped states $\ket{1,-1}$ and $\ket{2,1}$ using time-domain Ramsey interferometry.
The two states exhibit the same first-order Zeeman shift, thereby strongly reducing the sensitivity of the transition frequency to magnetic fields. 
For the  Ramsey measurements, the atoms are prepared in a coherent superposition driven by a pulsed microwave field $\omega_\text{extMW}$ from an external antenna and an additional radio frequency $\omega_\text{RF}$ fed to the Z-shaped trapping wire (green arrows in Fig.\,\ref{fig:Ramsey}a).
After a variable time $T_\text{Ramsey}$, a second MW + RF pulse is applied and the relative population in the two states is measured.
The populations in the two states oscillate with the difference  between the atomic frequency and the external frequency, $ \omega_\text{at}- (\omega_\text{extMW}+\omega_\text{RF})$.  
During the Ramsey sequence, the coplanar microwave cavity is driven by a field with a variable angular frequency $\omega_\text{dress}$ which is off-resonant to the atoms transition. 
This leads to an AC shift of the levels which depends on the detuning $\Delta$ between $\omega_\text{dress}$ and the atomic transition frequency. 
For a simple two-level system,  the off-resonant field shifts the atomic states by $\delta_\text{dress} = \pm\frac{\Omega_\text{dress}^2}{\Delta}$, where $\Omega_\text{dress}$ denotes the Rabi frequency of the dressing field and $\Delta = \omega_\text{dress} - \omega_0$ is the detuning between the dressing field and the atomic transition frequency. 
The plus (minus) sign is valid for the ground (excited) state. 
The level scheme of the atoms involving all relevant fields is depicted in Fig.\,\ref{fig:Ramsey}a.
For a microwave field which is linearly polarized perpendicular to the quantization axis, as it is in our case, the cavity field induces $\sigma^-$ and $\sigma^+$-transitions with equal field strength, as depicted by the red arrows.
This field hence couples the state $\ket{1,-1}$ to the states $\ket{2,-2}$ and $\ket{2,0}$. 
The state $\ket{2,1}$, on the other hand, is coupled to state $\ket{1,0}$.
This leads to a shift in the two-photon transition frequency $\ket{1,-1} \rightarrow \ket{2,1}$ by 
\begin{equation}
\delta_\text{dress} = - \Omega_\text{dress}^2\cdot\left( \frac{3}{\Delta_1} + \frac{1/2}{\Delta_2} + \frac{3/2}{\Delta_3} \right),
\label{eq:Omega2}
\end{equation}
which is measured in our experiment.
Here, $\Delta_i, i \in\{1,2,3\}$ denotes the detuning to the relevant atomic hyperfine transition. 
The numerical factors in the numerator are determined by the Clebsch-Gordan coefficients of the transitions.

For the measurement, the power of the microwave fed to the resonator and the magnetic offset field ($B_\text{off} = 0.315 \pm 0.003$\,mT) are held constant.

The measured frequency shift $\delta_\text{dress}$ in the Ramsey experiment is shown in Fig.\,\ref{fig:Ramsey}c.
As visible in the curve measured at $T=6.8$\,K, the dressing shift changes sign when the frequency of the dressing field is crossing an atomic resonance.
Variation of the dressing frequency affects the shift in two ways, via the detuning to the atomic transitions and via a change in the microwave intensity in the resonator.
Knowing the detuning to all involved levels, the normalized power of the microwave in the resonator, which is proportional to the square of the resonant Rabi frequency $\Omega_\text{dress}^2$, can be deduced from the dressing shift.
The calculated Rabi frequencies $\Omega_\text{dress}$ according to Eq.\,(\ref{eq:Omega2}) are shown as circles in Fig.\,\ref{fig:Ramsey}d.
The measurement was repeated for different temperatures of the superconducting chip, corresponding to different cavity resonance frequencies.
The result is compared with transmission spectra measured using a programmable network analyzer (solid lines in Fig.\,\ref{fig:Ramsey}d).
All curves are normalized to their maxima for the sake of comparability.
Lorentzian curves (dashed lines) fitted to the data points match the transmission spectra closely in centre frequency and peak width, which is on the order of $\kappa/2\pi \approx 2-3$\,MHz. 
We attribute deviations in the peak positions to uncertainties in the temperature regulation of the cryostat, which are of the order $\Delta T \approx 50$\,mK.

\subsection*{Coherent control of atomic states with cavity fields}
\begin{figure}
\centerline{\includegraphics[width=.5\textwidth]{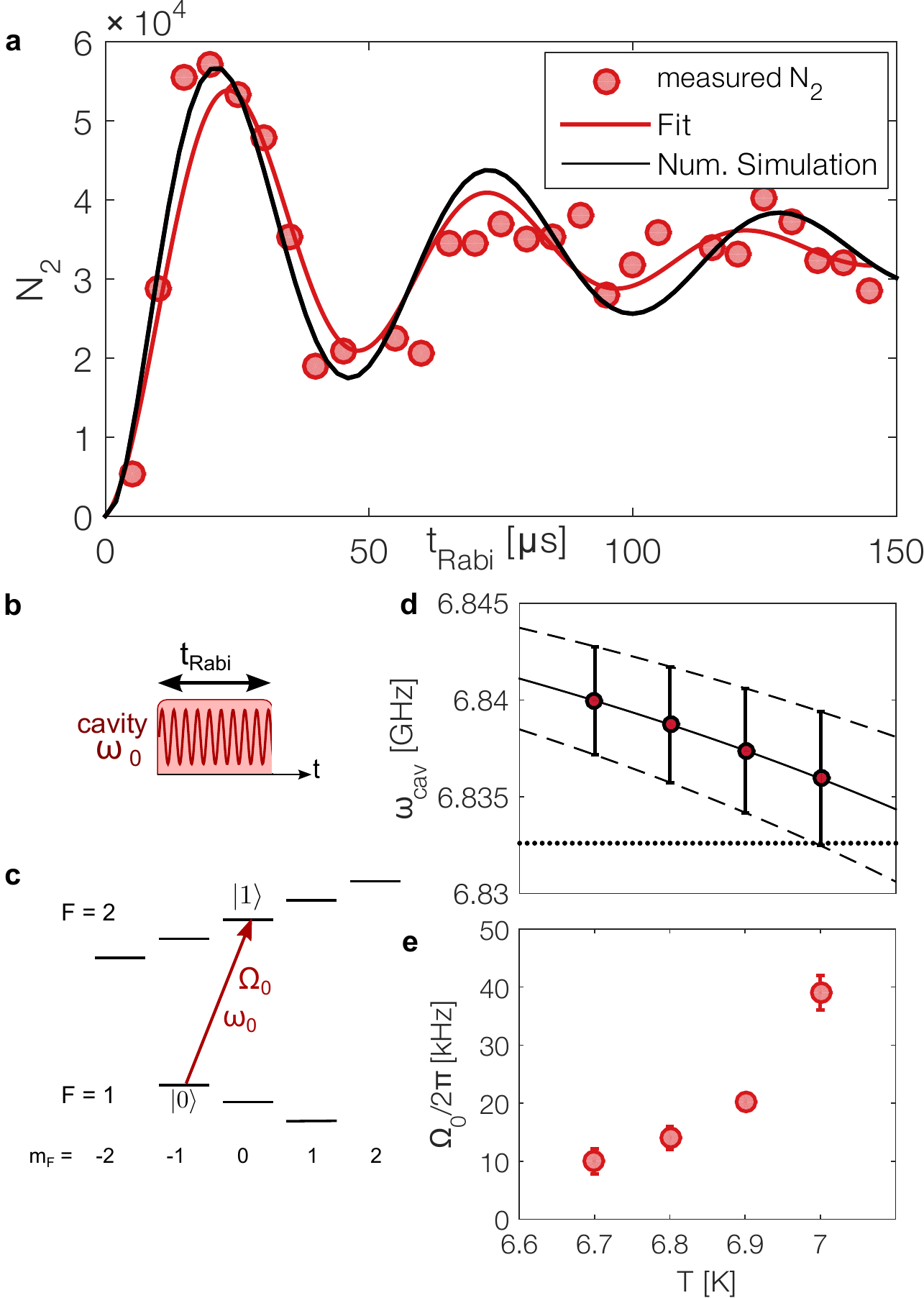}}
\caption{\textsf{\textbf{Cavity driven Rabi oscillations.} \textbf{a} Measurement of the atoms in state $\left|2,0 \right\rangle$ reveals resonant Rabi oscillations between $\left|1, -1 \right\rangle$ and $\left|2,0 \right\rangle$ for a cavity driving frequency of $\omega_0 = 2\pi\cdot6.83242$\,GHz. The chip temperature was set to $T = 6.9\,$K. The red solid line is a fit to the damped oscillation, the black line shows the result of the numerical simulations. \textbf{b} Timing sequence and \textbf{c} level scheme for the driven one-photon Rabi oscillations. \textbf{d} Temperature dependence of the cavity resonance frequency. The circles and bars indicate the peak and the width ($\pm \kappa $) of the cavity line  obtained from fits to the resonator transmission data. The solid and dashed lines indicate the fitted temperature dependence of the cavity frequency and linewidth (see Supplementary for details). The horizontal dotted line indicates the driving frequency, corresponding to the atomic resonance.  \textbf{e} Temperature dependence of the Rabi frequency. While the cavity is driven at the same frequency $\omega_0$ for all measurements, the temperature dependence of the cavity resonance leads to a change in the microwave intensity.}}
\label{fig:Rabi}
\end{figure}

When the electromagnetic cavity field is resonant with one of the (allowed) atomic transitions, the atoms undergo coherent Rabi oscillations between the ground and excited state.
The observation of these oscillations demonstrates coherent control over the internal atomic degrees of freedom. 
The Rabi frequency is given by  $\Omega_0 = \vec{\mu}\cdot\vec{B}_\text{MW}$, where $\vec{\mu}$ is the atomic magnetic moment and $\vec{B}_\text{MW}$ is the amplitude of the oscillating magnetic microwave field.
For the observation of these oscillations, we drive the cavity with a frequency $\omega_\text{0} = 2\pi\cdot6.83242$\,GHz, which is in resonance with the atomic transition $\ket{1,-1}\rightarrow\ket{2,0}$, but detuned roughly by twice the cavity linewidth $\kappa$ from the cavity resonance ($\omega_\text{cav} \approx 2\pi\cdot6.839$\,GHz) at a chip temperature $T = 6.9\,$K.
By state selective absorption imaging of the atoms, we observe resonant Rabi oscillations between the states $\ket{1,-1}$ and $\left|2,0 \right\rangle$ with a Rabi frequency $\Omega_\text{0} \approx 2\pi\cdot20$\,kHz (Fig.\,\ref{fig:Rabi}a).
By variation of the chip temperature between $T = 6.7$\,K and 7.0\,K, the cavity frequency is shifted with respect to the atomic transition (Fig.\,\ref{fig:Rabi}d).
This leads to a measurable change in the resonant Rabi frequency due to the altered MW power in the cavity, as visible in Fig.\,\ref{fig:Rabi}e.
Here, the Rabi frequency increases with higher temperatures, as the cavity frequency approaches the atomic transition frequency. 
For temperatures around $T = 7.2$\,K, the cavity resonance is shifted to coincide with the atomic resonance. 
However, at this temperature, the critical current of the superconducting coupling inductances is too low to support a stable magnetic trap.
We observe a damping in the single-photon Rabi oscillations with a time constant of $\tau \approx \SI{50}{\micro\second}$.
This damping is a result of the dephasing due to the inhomogeneous MW field of the cavity and the fact that Rabi oscillations are driven between two states with different magnetic moments. 
The magnetically trapped state $\ket{1,-1}$ is subjected to an energy shift of $\sim2\pi\hbar\cdot 7$MHz/mT, while the untrapped state $\ket{2,0}$ is in first order insensitive to magnetic fields. 
As a consequence, the resonance frequency between the two states is not uniform across the cloud and the atoms are only exactly on resonance at the centre of the trap.
A numerical simulation of a thermal cloud of $T_\text{at} = 2$\,\si{\micro\kelvin} trapped in a harmonic magnetic potential $\SI{20}{\micro \metre}$ above the cavity gap shows a damping time in excellent agreement with our measurement. 

In order to exploit the long coherence times of cold atoms, it is necessary to create superpositions between appropriate atomic states, which can both be trapped in the cavity. 
For $^{87}$Rb, such a state combination consists of the hyperfine levels $\ket{1,-1}$ and $\ket{2,1}$, which can both be trapped magnetically and exhibit excellent coherence properties. 
To this end, we start with an atomic cloud at a lower temperature of $T_\text{at} = 800\,$nK and $N_\text{at} \sim 3\times 10^4$ atoms in the state $\ket{1,-1}$.
\begin{figure}
\centerline{\includegraphics[width=.5\textwidth]{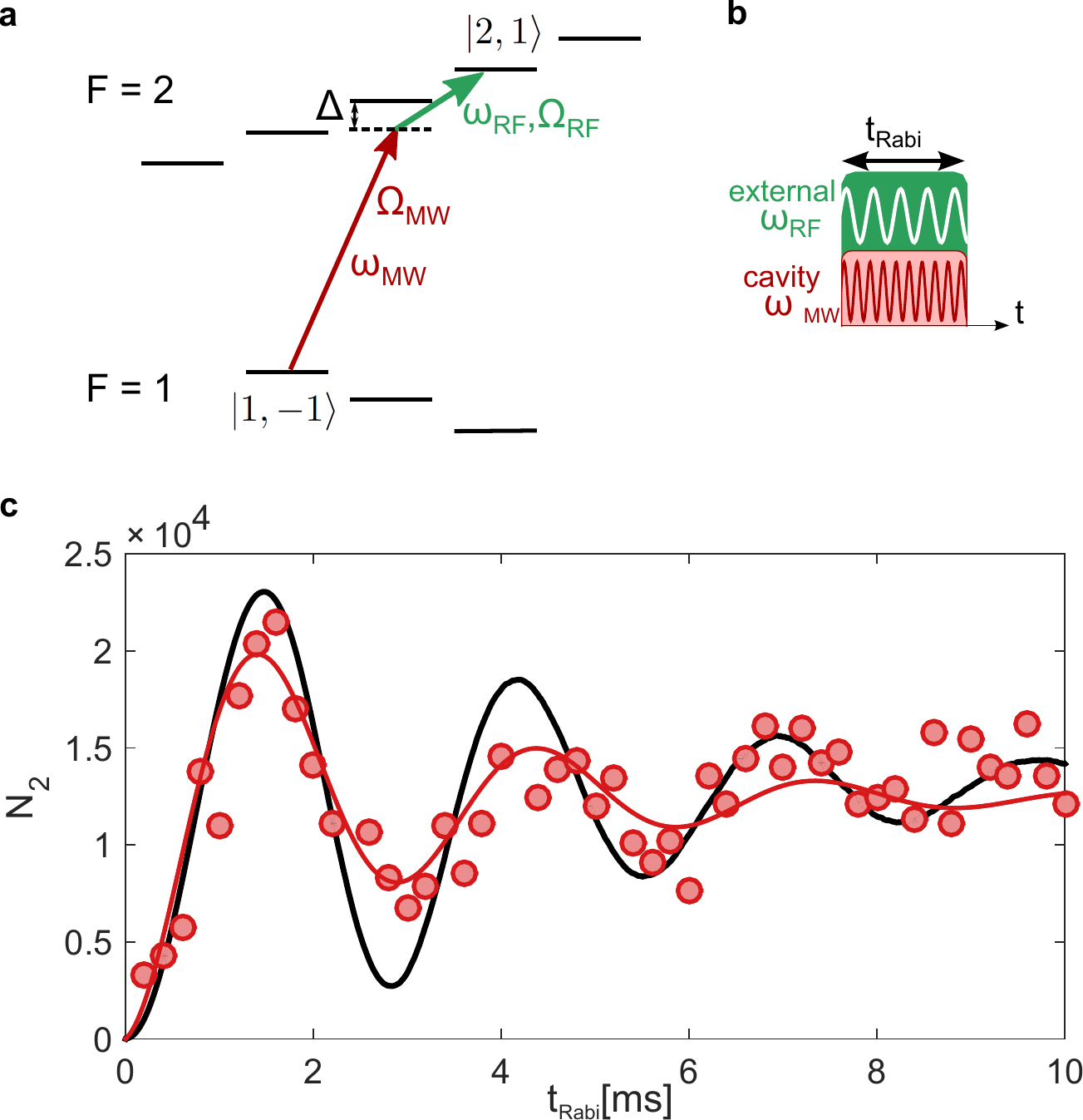}}
\caption{\textsf{\textbf{Two-photon Rabi oscillations.} \textbf{a}  Level scheme and \textbf{b} timing sequence for the two photon Rabi oscillations between the trapped states $\left| 1, -1 \right\rangle$ and $\left|2,1 \right\rangle$. } \textbf{c} Observation of two-photon Rabi oscillations between states $\ket{1,-1}$ and $\ket{2,1}$ (circles) and fit of the damped oscillation (red), yielding a damping time of $\tau = 5$\,ms due to to the inhomogeneity of the MW field amplitude across the cloud. The black solid line shows a numerical simulation of the state evolution for an ensemble of thermal atoms moving in the trap. }
\label{fig:Rabi2}
\end{figure}
In order to prepare a coherent superposition of the two states, we drive the cavity with the MW field $\omega_\text{MW}$ and employ an additional external RF field $\omega_\text{RF}$, with a detuning of $\Delta = 2\pi\cdot300$\,kHz to the intermediate state $\ket{2,0}$ (cf. Fig\,\ref{fig:Rabi2}a).
If the two corresponding Rabi frequencies are small compared to the intermediate detuning, i.e.\,$\Omega_\text{MW}, \Omega_\text{RF} \ll \Delta$, the population of the intermediate state can be neglected. 
In this case, the two-photon Rabi frequency $\Omega_\text{2Ph}$ can be calculated by adiabatic elimination of the intermediate state $\Omega_\text{2Ph} = \Omega_\text{MW}\Omega_\text{RF}/\Delta$ \cite{Gentile1989}.
By driving the two fields with variable pulse lengths, we observe Rabi oscillations with $\Omega_\text{2Ph} = 2\pi\cdot340$\,Hz, and a dephasing on the order of $\tau \sim 5$\,ms (Fig.\,\ref{fig:Rabi2}c).
A numerical simulation of an ensemble of non-interacting atoms in a magnetic trap reveals damping on the same timescale.
As in the one-photon case, the dephasing is mainly due to the variation of the microwave field strength over the size of the atomic cloud (see Supplementary section S5).

\subsection*{Discussion}
\vspace{-2mm}
To make the presented cold atom-superconductor hybrid device a useful, high-coherence quantum resource, several aspects need to be addressed and optimized. 
In particular, dephasing during the Rabi pulses should be reduced and the coupling between atoms and the cavity increased. 

Dephasing due to inhomogeneous coupling can, as seen in the experiment above, be a limitation for the high-fidelity creation of superposition states needed in information processing. 
The inhomogeneity seen by the atomic ensemble can be reduced by reducing the cloud temperature, yielding smaller cloud extension in the trap (see Supplementary section S5). 
Several experiments have furthermore shown that reliable superpositions or quantum gates can be achieved in spite of this temporal or spatial variation of Rabi frequencies, as the related dephasing can be overcome using more elaborate MW and RF pulses using optimal control theory \cite{Schulte-Herbrueggen2011,Dolde2014}. 
For our geometry, we have estimated the coupling between a single atom and a single cavity photon to be $g \approx 2\pi\cdot 0.5$\,Hz.
Various means can be used to increase the coupling strength between the atoms and the cavity field.
By decreasing the width of the gap $W$ between the centre conductor and ground planes of the cavity, the magnetic field per photon could be increased according to $B_\text{ph} \propto 1/W^2$, but would require the atoms to be trapped closer to the chip surface.
By changing the resonator layout from CPW to lumped element resonator, the inductance and dimensions of the resonator could be decreased, leading to a significant enhancement of the current per photon and hence magnetic field $B_\text{ph}$.
Finally, the electric field of the cavity mode could be used to couple neighbouring Rydberg states, exploiting the large electric dipole moments of Rydberg states \cite{Yu2016}. 
A conservative estimate of the coupling between Rydberg atoms and the field of a higher harmonic mode for our geometry yields a coupling strength on the order of 1-2\,MHz, which is on the same order as the cavity linewidth.

\section*{Conclusion}
\vspace{-2mm}
In summary, we have experimentally demonstrated coupling of ultracold ground state atoms to a driven superconducting coplanar waveguide resonator.
Coupling was shown both in resonant Rabi oscillation and in dressing the frequency of an atomic clock state pair. 
Future measurements will explore collective effects of cold atoms to the cavity mode and work towards strong coupling between the superconducting resonator and Rydberg atoms.
These experiments are the first step towards the implementation of cold atoms as a quantum resource in a hybrid quantum architecture.
\section*{Methods}
\footnotesize{

\subsection*{Atomic cloud preparation}
\vspace{-2mm}
The atomic ensemble is prepared in a room-temperature setup and transported to a position below the superconducting atom chip using an optical dipole trap that is moved using a lens mounted on an air-bearing translation stage (cf.\,Ref.\,\cite{Cano2011} for details).
Atoms are subsequently trapped in a magnetic trap generated by currents in the Z-shaped Nb wire and an external homogeneous bias field.
The Z-wire configuration leads to a Ioffe-Pritchard-type magnetic microtrap with a non-zero offset field $B_\text{off}$ at the trap minimum.
We load $\sim 10^6$ atoms at a temperature of $\sim \SI{1}{\micro \kelvin}$ into the magnetic chip trap.
After adiabatic compression, the cloud is transferred into the mode volume of the resonator by rotating the external bias field and switching off the current in the Z-trap.
Screening currents in the resonator, which conserve the flux in the closed superconducting loop, lead to the formation of a magnetic trap with oscillation frequencies $\omega_x = 2\pi\cdot 400$\,s$^{-1}$, $\omega_y = 2\pi\cdot 25$\,s$^{-1}$, $\omega_z = 2\pi\cdot 600$\,s$^{-1}$  below the gap of the waveguide cavity, $\SI{20}{\micro \meter}$ from the chip surface.
During the transfer into the tight trap, the atomic cloud is heated up to a temperature of $T_\text{at} \sim \SI{2}{\micro \kelvin}$.
At the cavity position, we perform radiofrequency evaporation to further cool the atomic ensemble.

\subsection*{Experimental cycle and state selective detection}
In order to measure the atomic state, the following experimental cycle is repeated every $\sim26$\,s.
After preparation of an atomic cloud, transporting it to the superconducting chip and loading into the cavity, as described above, all atoms are in the hyperfine state $\ket{1,-1}$.
Subsequently, we apply one MW (+RF) pulse of variable length $t_\text{Rabi}$ for the measurement of Rabi oscillations, or two $\pi/2$-pulses of fixed length with a variable hold time $t_\text{Ramsey}$ in between for the Ramsey interferometry sequence.
At the end of the sequence, we can measure the number of atoms in both of the states.
First, the number of atoms in F=2 is measured by illuminating the cloud with light resonant to the $5S_{1/2}, F{=}2 \rightarrow 5P_{3/2}, F{=}3$ transition. 
The shadow of the atoms is imaged on a CCD camera and the measured optical density is used to determine the atom number. 
We then pump the atoms from $F=1$ into $F=2$ by illumination with a laser resonant with the $5S_{1/2}, F{=}1 \rightarrow 5P_{3/2}, F{=}2$ transition. 
From the $5P_{3/2}, F{=}2$ state, the atoms decay into $5P_{1/2}, F{=}2$ in $\sim 30$\,ns and the atoms are imaged on a second CCD camera as described above. 
\normalsize

\subsection*{Acknowledgements}
\vspace{-2mm}
This work was supported by the Deutsche Forschungsgemeinschaft (SFB TRR21) and the European Commission (FP7 STREP project ``HAIRS'').
H.H. and B.F. acknowledge additional support from the  Carl Zeiss Stiftung and the Research Seed Capital (RiSC) programme of the MWK Baden-W\"urttemberg. 
\subsection*{Author contributions}
\vspace{-2mm}
D.K., R.K., J.F., and H.H. designed and mounted the experiment. D.B., D.W., B.F. and H.H. developed and fabricated the superconducting chip.  H.H. and L.Y.L. carried out the experiments, H.H.,D.B., L.Y.L. and B.F. analyzed the data. H.H., D.B. and B.F. performed the numerical simulations. L.S. provided the microwave dressing theory. D.K., R.K. and J.F. supervised the project. H.H., D.B. and J.F. edited the manuscript. All authors discussed the results and contributed to the manuscript. 
\subsection*{Competing financial interest}
\vspace{-2mm}
The authors declare no competing financial interests.

\pagebreak
\widetext

\noindent\textbf{\textsf{\Large Supplementary Materials: Coupling ultracold atoms to a superconducting coplanar waveguide resonator}}

\normalsize
\vspace{.3cm}

\noindent\textsf{H. Hattermann, D. Bothner, L. Y. Ley, B. Ferdinand, D. Wiedmaier, L. S\'ark\'any, R. Kleiner, D. Koelle, J. Fort\'agh}

\vspace{.2cm}
\noindent\textit{CQ Center for Quantum Science in LISA$^+$, Physikalisches Institut, Eberhard Karls Universit\"at T\"ubingen, 	
Auf der Morgenstelle 14, D-72076 T\"ubingen, Germany}

\renewcommand{\thefigure}{S\arabic{figure}}
\renewcommand{\theequation}{S\arabic{equation}}

\renewcommand{\thesection}{S\arabic{section}}
\renewcommand{\bibnumfmt}[1]{[S#1]}

\setcounter{figure}{0}
\setcounter{equation}{0}

\section{Atom chip design and fabrication}

Our atom chip combines two structures, a $Y = 100\,\mu$m wide Z-shaped superconducting Nb strip  for the application of directed and low frequency currents as well as a superconducting coplanar waveguide resonator with a resonance frequency of $\omega_\mathrm{Res} \approx 2\pi\cdot6.85\,$GHz, near-resonant with the ground state hyperfine transition frequency of $^{87}$Rb atoms.
All structures are patterned onto a $h_\mathrm{S} = 330\,\mu$m thick sapphire substrate by means of optical lithography, thin film deposition and microfabrication .
A schematic of the atom trapping region on the chip is shown in Fig.~\ref{fig:Supp_Par}a and a cross-sectional view along the dotted line in \ref{fig:Supp_Par}a is shown in \ref{fig:Supp_Par}b.
The full chip layout is shown in Fig.~1a of the main paper.
The coplanar microwave resonator has a centre conductor width of $S = 30\,\mu$m and two ground planes, which are separated from the centre conductor by a gap of $W = 16\,\mu$m, targeting a characteristic impedance $Z_0 = 50\,\Omega$.
In order to facilitate the magnetic trapping of atoms closely above the gaps of the waveguide structure, the magnetic field distorting superconducting ground planes had to be removed partially.
As we observe strong parasitic resonances when parts of the ground planes are missing (probably due to a parasitic mutual inductance between the trapping wire and the waveguide structure and due to the excitation of chip resonances), we substituted the removed superconducting parts by a normal-conducting Au metallization layer, restoring a good ground connection along the whole resonator.
Thus, the trapping wire is embedded into one of the ground planes and galvanically connected to all metallization parts on the chip.
As superconductor we use niobium, and as normal conductor we use gold on top of a thin adhesion layer of titanium.
The thicknesses of the three films are $h_\mathrm{Nb} = 500\,$nm, $h_\mathrm{Au} = 400\,$nm, and $h_\mathrm{Ti} = 4\,$nm, cf. Fig.~\ref{fig:Supp_Par}b.
Between the superconducting parts and the normal-conducting parts, there is a $O = 10\,\mu$m wide overlap region, ensuring a low contact resistance.
In order to minimize additional microwave losses induced by the presence of the normal conductor, we only replaced the superconductor by gold in the trapping region ($\sim 15\%$ of the total resonator length) and kept also a $G = 50\,\mu$m part of the ground plane in this region superconducting.
The normal conducting region in between this remaining superconducting part of the ground plane and the superconducting trapping wire is $D+2O = 120\,\mu$m wide, cf. Fig.~\ref{fig:Supp_Par}b.
\begin{figure}[h]
	\centering {\includegraphics[scale=0.7]{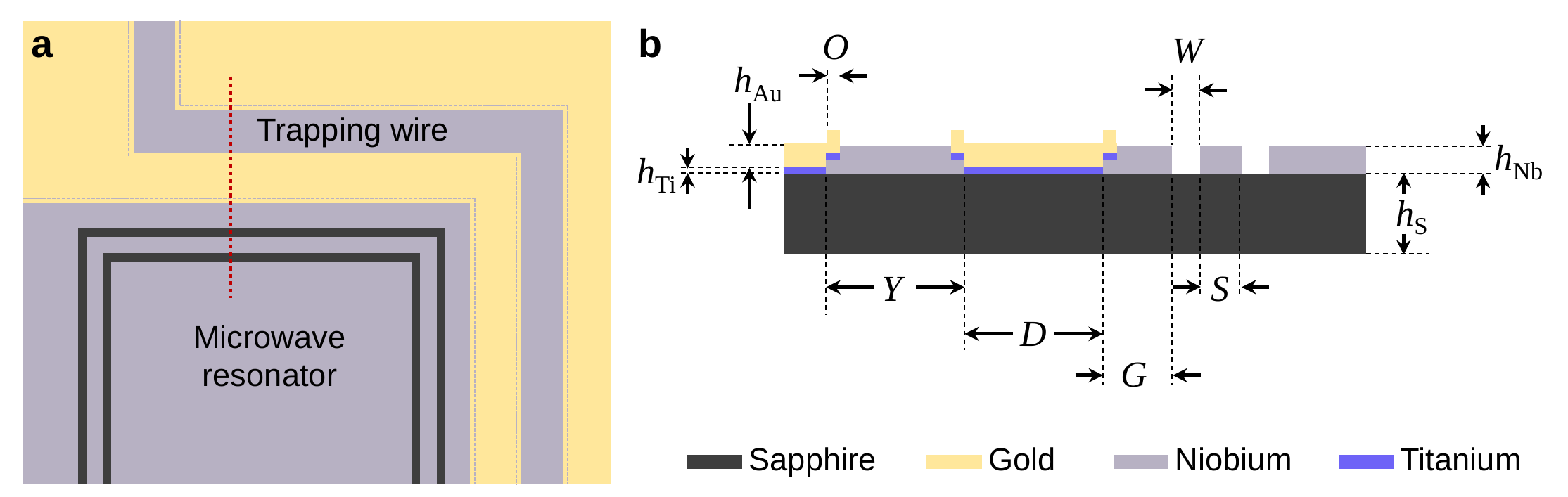}}
	\caption{\textsf{\textbf{Atom chip layout and parameters.} \textbf{a} Schematic top view of the trapping region of the atom chip. A Z-shaped atom trapping wire passes by a coplanar microwave resonator structure. The trapping wire and the core region of the microwave resonator consist of superconducting Nb, the two structures are galvanically connected by a normal conducting gold layer in order to guarantee well-defined microwave properties. \textbf{b} Cross section along the red dotted line in \ref{fig:Supp_Par}a, depicting and defining all relevant materials, thicknesses and geometrical parameters of the device. Thicknesses are not to scale.}}
	\label{fig:Supp_Par}
\end{figure}
The device fabrication is schematically shown in Fig.~\ref{fig:Supp_Fab}.
It starts with the DC magnetron sputtering of the Nb onto a bare r-cut Sapphire substrate.
By means of optical lithography and SF$_6$ reactive ion etching, we pattern the superconducting parts.
Next, we cover most of the superconducting parts -- except for the $10\,\mu$m wide overlap region -- with photoresist and deposit the normal conducting metal on top.
To do so, we first remove $~200\,$nm of the Nb in the overlap region by another SF$_6$ reactive ion etching step in order to get rid of photoresist residues and a possible native oxide layer on top of the Nb and in addition to reduce the substrate-Nb step height. 
Then, we in-situ deposited the Ti adhesion layer by means of electron beam evaporation and the Au layer by DC magnetron sputtering.
We finalized the fabrication by lifting off the normal conducting parts in hot acetone supported by ultrasound.
\begin{figure}[h]
	\centering {\includegraphics[scale=0.7]{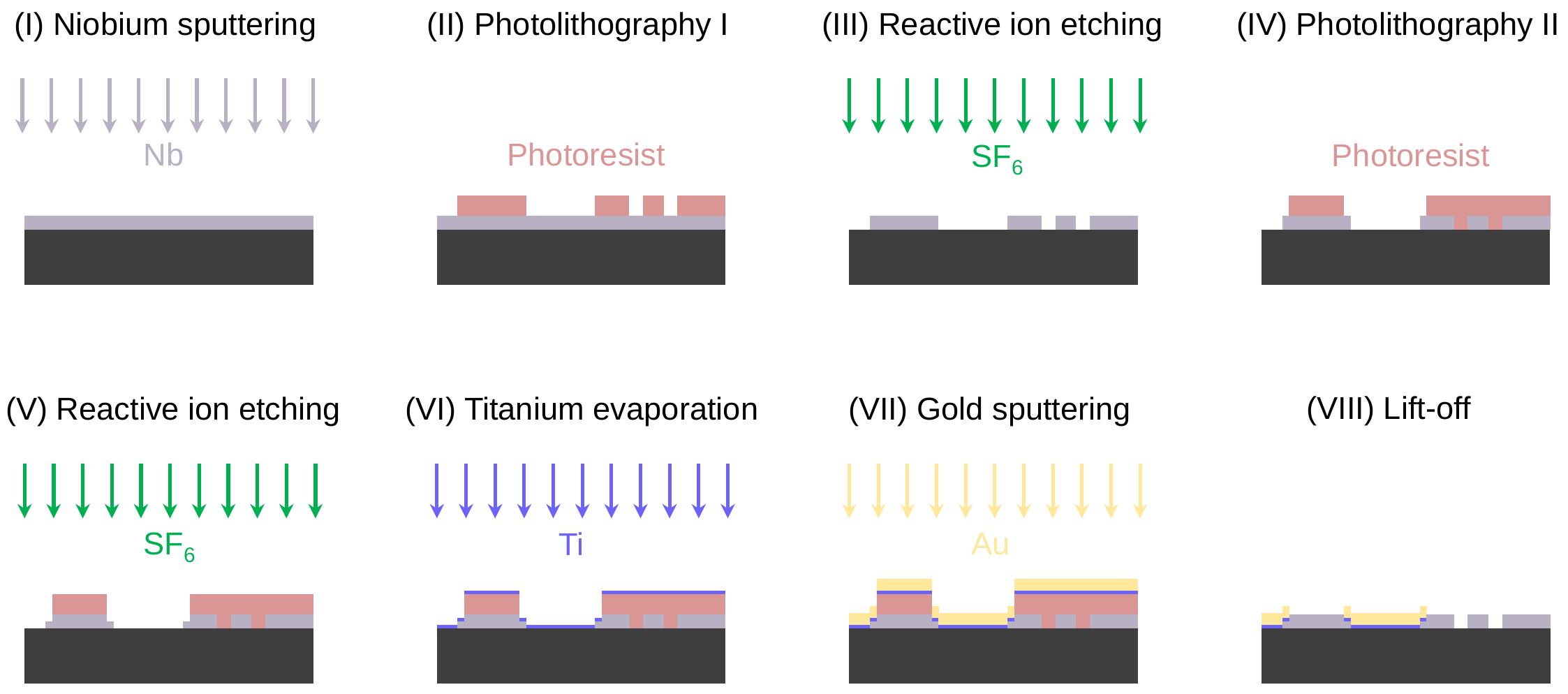}}
	\caption{\textsf{\textbf{Atom chip fabrication}. Schematic fabrication sequence of the chip used in this experiment. Thicknesses are not to scale. (I) DC magnetron sputtering of Nb onto a Sapphire substrate. (II), (III) Photolithopgraphy and reactive ion etching defining the superconducting chip parts. (IV) Protection of the superconducting parts with photoresist, except for a $O = 10\,\mu$m wide overlap edge region. (V) Removal of the native oxide in the overlap region and reduction of the substrate-Nb step height by reactive ion etching. (VI) Electron beam evaporation of a titanium adhesion layer. (VII) DC magnetron sputtering of Au. Steps (V)-(VII) are performed in-situ. (VIII) Ultrasound assisted lift-off of Au/Ti in warm acetone.}}
	\label{fig:Supp_Fab}
\end{figure}

\section{Cavity parameters}

The microwave resonator used in this experiment is a half wavelength ($\lambda/2$) transmission line cavity based on a coplanar waveguide with charactersitic impedance $Z_0 \approx 50\,\Omega$ and attenuation constant $\alpha$.
The transmission line cavity has a length $l_0 \approx 9.3\,$mm and a fundamental mode resonance frequency $\omega_\mathrm{Res} = 2\pi\cdot 6.85\,$GHz at a temperature of $\sim 5\,$K.
Around its resonance frequency, the waveguide resonator can be modelled as an inductively coupled series RLC circuit, cf. Fig.~\ref{fig:Supp_Cav}a and \ref{fig:Supp_Cav}b with the equivalent lumped element resistance $R$, inductance $L$ and capacity $R$ \cite{Bothner13}:

\begin{equation}
R = Z_0\alpha l_0, ~~~~~ L = \frac{\pi Z_0}{2\tilde{\omega}_\mathrm{Res}}, ~~~~~ C = \frac{2}{\pi \tilde{\omega}_\mathrm{Res} Z_0}
\end{equation}
where $\alpha$ is the attenuation constant of the coplanar waveguide and $\tilde{\omega}_\mathrm{Res}$ is the "uncoupled" resonance frequency, i.e., the resonance frequency corresponding only to the electrical length of the cavity.

\begin{figure}[h]
	\centering {\includegraphics[scale=0.75]{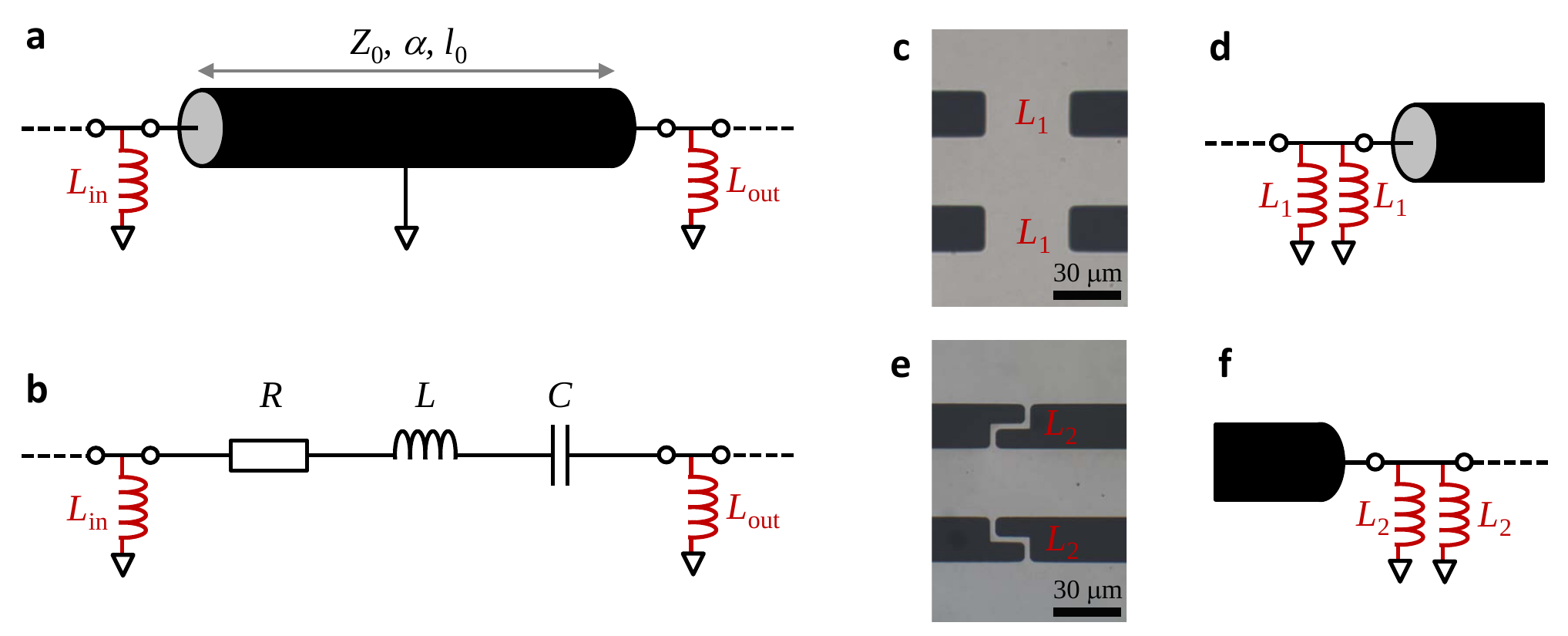}}
	\caption{\textsf{\textbf{Resonator parameters and description}. \textbf{a} Schematic of an inductively coupled transmission line cavity as used in this experiment. The transmission line resonator is characterized by its characteristic impedance $Z_0$, its length $l_0$ and its attenuation constant $\alpha$. The cavity is coupled at both ends to transmission feedlines via shunt inductors $L_\mathrm{in}$ and $L_\mathrm{out}$. \textbf{b} Lumped element circuit equivalent of \ref{fig:Supp_Cav}a with the equivalent resistor $R$, the equivalent inductor $L$ and the equivalent capacitor $C$. \textbf{c} [\textbf{e}] shows an optical image of the input [output] coupling inductors of our device and \textbf{d} [\textbf{f}] shows its circuits equivalent. As in the coplanar waveguide geometry we have two parallel shunt inductors $L_1$ [$L_2$] to ground at the input [output] port, the total input [output] coupling inductance is given by $L_\mathrm{in} = L_1/2$ [$L_\mathrm{out} = L_2/2$].}}
	\label{fig:Supp_Cav}
\end{figure}

For driving the resonator and reading out its frequency dependent response, the cavity is weakly coupled to two feedlines by shunt inductors between the centre conductor and the ground planes at both ends, cf. Fig.~1 of the main paper.
The shunt inductors at the input port are shown in Fig.~\ref{fig:Supp_Cav}c.
Each of the two superconducting shunts to ground is $36\,\mu$m wide and $16\,\mu$m long.
With the software package 3D-MLSI \cite{Khapaev03}, we determined each of the two shunt inductances to be $L_1 = 2.94\,$pH, giving a total input port coupling inductance $L_\mathrm{in} = L_1/2 = 1.47\,$pH.
At the output port, cf. Fig.~\ref{fig:Supp_Cav}e, the shunt inductors are $4\,\mu$m wide and $30\,\mu$m long, giving an inductance per shunt of $L_2 = 12.88\,$pH.
Thus, the total inductance at the output port is $L_\mathrm{out} = L_2/2 = 6.44\,$pH.
For $\tilde{\omega}_\mathrm{Res} L_\mathrm{in}, \tilde{\omega}_\mathrm{Res} L_\mathrm{out} \ll Z_0$ the resonance frequency of the coupled circuit is shifted due to the coupling inductors according to

\begin{equation}
\omega_\mathrm{Res} = \frac{1}{\sqrt{(L+L_\mathrm{in} + L_\mathrm{out})C}}.
\end{equation}

The external linewidth of the resonator due to losses through the input port is given by \cite{Bothner13}

\begin{equation}
\kappa_\mathrm{ex1} = \omega_\mathrm{Res}\frac{\pi}{2} \frac{L_\mathrm{in}^2}{L^2}\approx 2\pi\cdot 7\,\mathrm{kHz}.
\end{equation}

For the output port, we find

\begin{equation}
\kappa_\mathrm{ex2} = \omega_\mathrm{Res}\frac{\pi}{2} \frac{L_\mathrm{out}^2}{L^2} \approx 2\pi\cdot 134\,\mathrm{kHz}.
\end{equation}

These linewidths correspond to a total external linewidth

\begin{equation}
\kappa_\mathrm{ex} = 2\pi\cdot 141\,\mathrm{kHz}
\end{equation}

or a total external quality factor

\begin{equation}
Q_\mathrm{ex} = \frac{\omega_\mathrm{Res}}{\kappa_\mathrm{ex}} \approx 5\cdot 10^4.
\end{equation}

In liquid helium, at temperature $T_s = 4.2\,$K, we measure a total quality factor of $Q\approx 10000$, indicating that the majority of the losses is due to thermal quasiparticles in the superconductor as well as due to dissipation in the normal conducting parts and the interfaces between the different metals.

\section{Cavity temperature dependence}

\subsection{Temperature calibration}

The magnetic penetration depth $\lambda_\mathrm{L}$ in a BCS superconductor shows a temperature dependence, which can be approximately captured by \cite{Tinkham2004}

\begin{equation}
\lambda_\mathrm{L}(T) = \frac{\lambda_\mathrm{L}(T=0)}{\sqrt{1-\left(\frac{T_s}{T_c}\right)^4}}
\end{equation}
with the sample temperature $T_s$ and the superconducting transition temperature $T_c$. 
The origin of this temperature dependence is the temperature dependence of the superconducting charge carrier density.
The total inductance of a superconducting resonator is given by the sum of the temperature independent geometric inductance $L_g$ and the kinetic inductance, $L_k(T)$, which takes into account the kinetic energy of the superconducting charge carriers.
For superconductors with a thickness larger than twice the penetration depth, the kinetic inductance is related to the magnetic penetration depth via

\begin{equation}
L_k(T) = \chi_g\mu_0\lambda_\mathrm{L}(T),
\end{equation}
where $\chi_g$ is a geometrical factor, taking into account the spatial distribution of the superconducting current density.
In our samples, we have $h_\mathrm{Nb} = 500\,$nm and typically $\lambda_\mathrm{L}(T = 0) \sim 100\,$nm.
Thus, up to $T_s/T_c \approx 0.95$, which is much larger than all values of $T_s/T_c$ in our experiment,  $h_\mathrm{Nb} > 2\lambda_\mathrm{T}$ is fulfilled.
In general, also the coupling inductors have a kinetic contribution, but due to $L \gg L_\mathrm{in}, L_\mathrm{out}$ in our device, we neglect this small correction here.
With the temperature dependent kinetic inductance, the resonance frequency is given by

\begin{equation}
\omega_\mathrm{Res}(T) = \frac{\omega_\mathrm{Res0}}{\sqrt{1 + \frac{L_k(T)}{L_0}}},
\label{eqn:wres_Tdep}
\end{equation}
where $L_0 = L_g + L_\mathrm{in} + L_\mathrm{out}$ is the inductance of the cavity without the kinetic contribution and $\omega_\mathrm{Res0}=1/\sqrt{L_0C}$ is the resonance frequency for $L_k = 0$ (not for $T = 0$).
\begin{figure}[h]
	\centering {\includegraphics[scale=0.75]{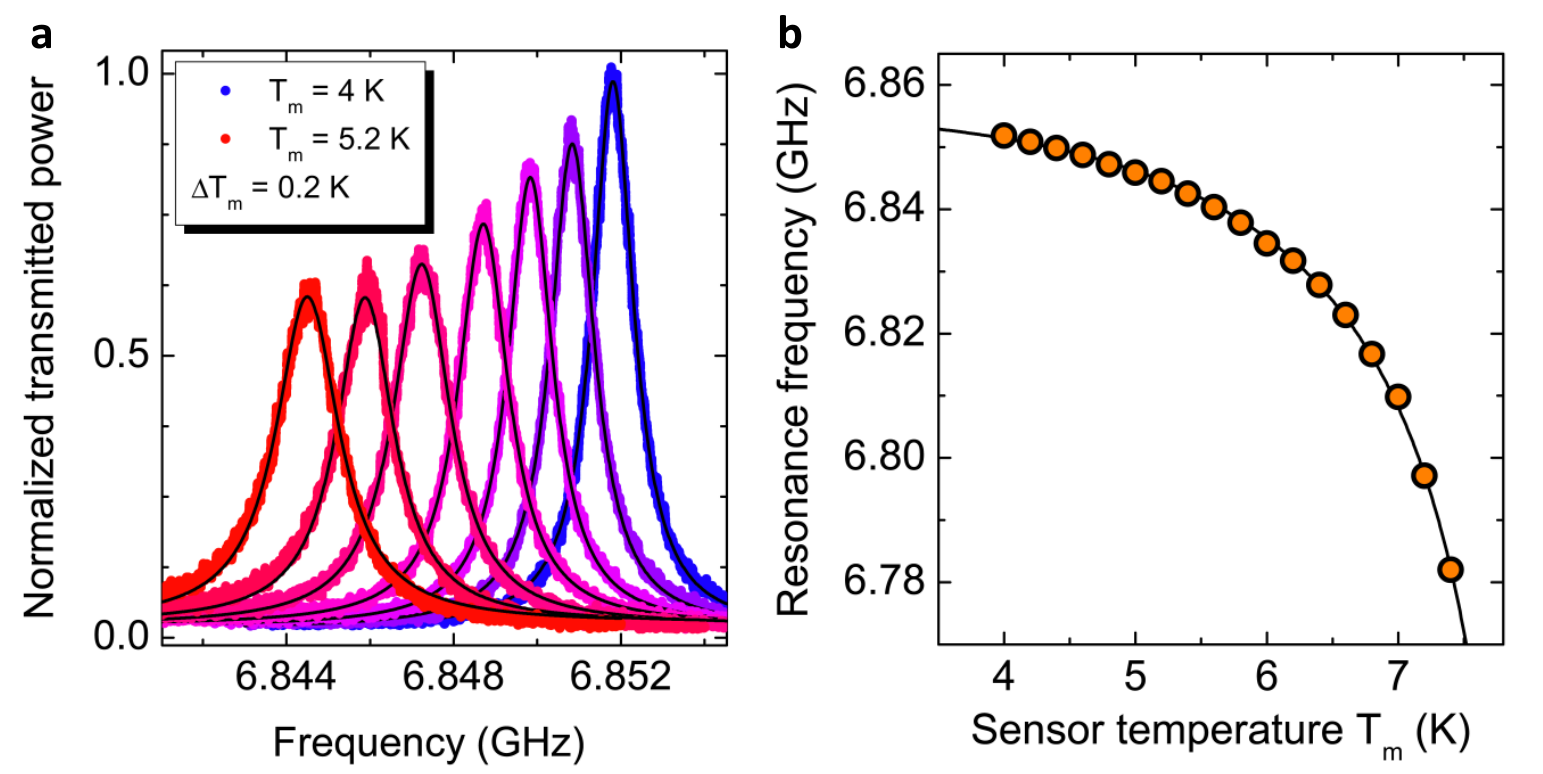}}
	\caption{\textsf{\textbf{Temperature calibration}. \textbf{a} Cavity transmission spectra measured for sensor temperatures $4\,\mathrm{K}\leq T_m \leq 5.2\,\mathrm{K}$ in steps of $\Delta T_m = 0.2\,$K. With increasing temperature, the resonance frequency shifts to lower values. Black lines are Lorentzian fits. \textbf{b} Cavity resonance frequency $\omega_\mathrm{Res}/2\pi$ vs sensor temperature. Circles are data extracted from the measurements and the black line is an analytical approximation curve (for details see text).}}
	\label{fig:Supp_Tdep}
\end{figure}

In our experiment, we take advantage of the temperature dependence of the cavity resonance frequency to tune it close to the atomic transition frequency.
Figure~\ref{fig:Supp_Tdep}a shows (smoothed) transmission spectra for different temperatures measured with the sensor mounted to the helium flow cryostat, which also hosts the chip.
We observe the resonance frequency shifting towards lower values with increasing temperature.
In Fig.~\ref{fig:Supp_Tdep}b, we plot the extracted resonance frequency vs the measured temperature $T_m$.
As the thermometer is positioned inside the coldfinger of the flow cryostat $\sim10\,$cm from the chip itself, we expect the sample temperature $T_s$ to be different from the sensor temperature $T_m$ by an offset temperature $T_\mathrm{off}$, i.e.,

\begin{equation}
T_s = T_m + T_\mathrm{off}.
\label{eqn:Ts}
\end{equation}

We note that we use a calibrated sensor and thus that the offset is not related to uncertainty of the sensor measurement, but due to the nature of the setup \cite{Cano10}.
The chip and the microwave amplifier are mounted on a $10\,$cm high sample holder of oxygen-free high-conductivity copper.
The cooling power of the chip is mainly limited by the thermal conductivity through the interfaces between the cryostat and the chip holder and between chip holder and the sapphire chip.
Due to the requirement to have optical access to the chip region, $5\,$mm high slits have been cut into the thermal shield at $20\,$K, which encloses the coldfinger tip and the sample holder in order to minimize the thermal radiation from the room temperature environment.
The final temperature of the chip is given by a combination of the cooling power from the coldfinger and the heating power due to thermal radiation from the environment.
We find a very good agreement between the experimentally determined resonance frequencies shown in Fig.~\ref{fig:Supp_Tdep}b, the transition temperature of our Nb $T_c = 9.2\,$K and Eq.~(\ref{eqn:wres_Tdep}) when we assume $T_\mathrm{off} = 1.05\,$K, $\omega_\mathrm{Res0} = 2\pi\cdot6.94378\,$GHz and a kinetic inductance participation ratio $L_k(T=0)/L_0 = 0.02589$.
The result is shown as black line in Fig.~\ref{fig:Supp_Tdep}b and gives us a rough estimate for the temperature offset between sample and sensor.

\subsection{Temperature fine calibration and full cavity characterization}

As the offset temperature $T_\mathrm{off}$ is not exactly constant between 5\,K and 9\,K and as all our experiments are done within a limited temperature window of $\sim 1\,$K, we performed a more detailed cavity characterization in the corresonding temperature interval.
The results of this detailed cavity characterization are shown in Fig.~\ref{fig:Cavity_Zoom}.
In Fig.~\ref{fig:Cavity_Zoom}a, we plot the resonance frequency vs the sample temperature, where the sample temperature was determined from the analytical approximation shown as black line.
To achieve the best match in this temperature region, we had to adjust the offset temperature to $T_\mathrm{off} = 1.09\,$K, but kept all other parameters used above.

\begin{figure}[h]
	\centering {\includegraphics[scale=0.75]{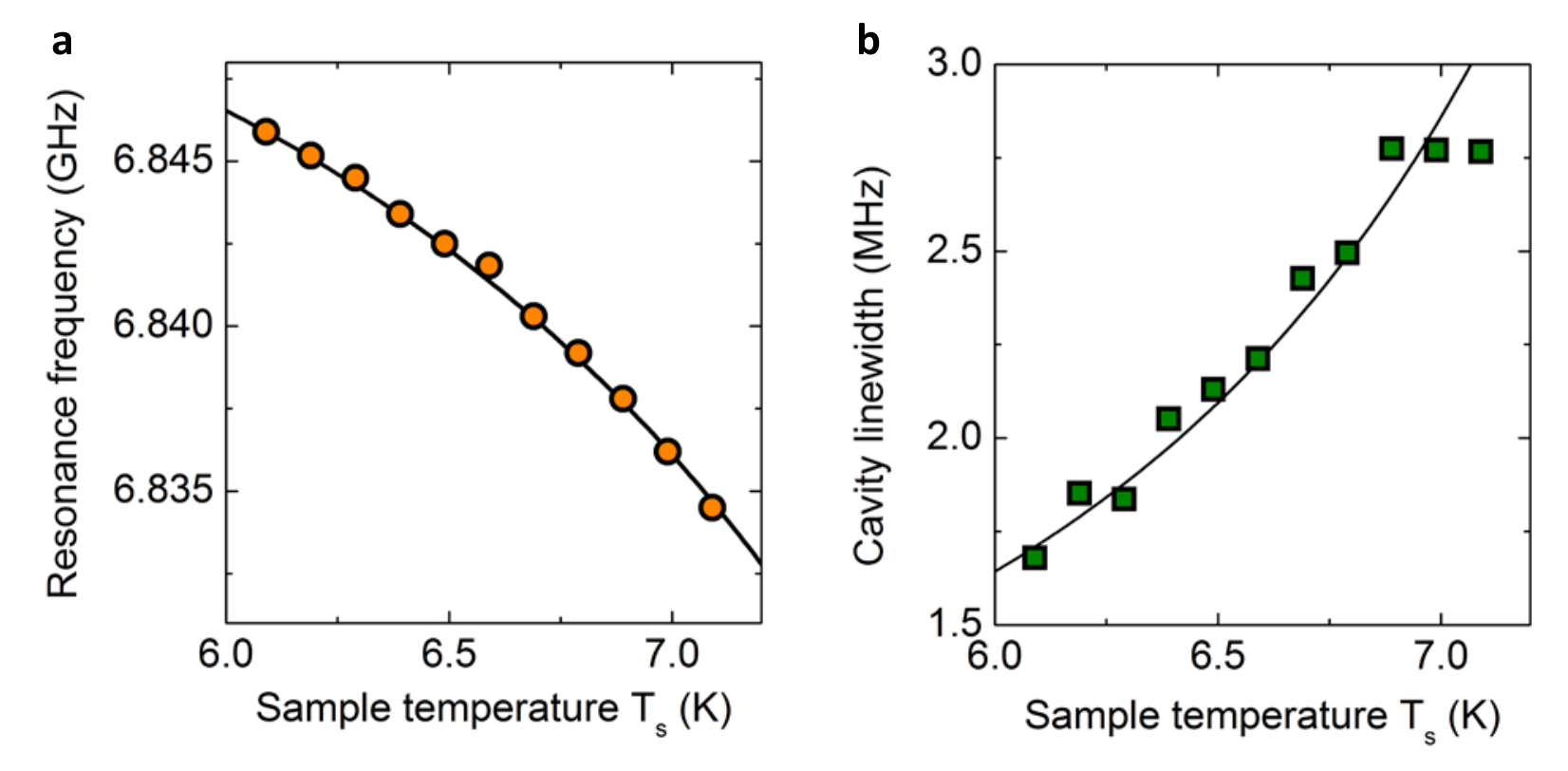}}
	\caption{\textsf{\textbf{Temperature dependence of the cavity parameters}. \textbf{a} Cavity resonance frequency $\omega_\mathrm{Res}/2\pi$ vs sample temperature. Circles are data extracted from the measurements and the black line is an analytical approximation curve (details see text). \textbf{b} Cavity linewidth $\kappa/2\pi$ vs sample temperature extracted from Lorentzian fits. Squares are experimental data and the black line is an approximation based on the two-fluid model (details see text). The data point at $6.99\,$K is linearly interpolated from points at $6.89\,$K and $7.09\,$K.}}
	\label{fig:Cavity_Zoom}
\end{figure}

In addition to the resonance frequency, we also extracted the resonance linewidth $\kappa$ for each temperature, which is shown in Fig.~\ref{fig:Cavity_Zoom}b.
From the two-fluid model \cite{Buckel2004, Tinkham2004}, it follows that the surface resistance of a superconductor is given by

\begin{equation}
R_s = \frac{1}{2}\omega^2\mu_0^2\sigma_1\lambda_\mathrm{L}^3
\end{equation}
where $\sigma_1 \propto n_n/n_e$ is the real part of the complex two-fluid conductivity with the quasiparticle density $n_n$ and the total electron density $n_e$.
From the temperature dependence of $\lambda_\mathrm{L}$ and the two-fluid model, the temperature dependence of the superconducting charge carrier density is given by

\begin{equation}
\frac{n_s(T)}{n_e} = 1-\left(\frac{T_s}{T_c}\right)^4.
\end{equation}

This leads to the quasiparticle density fraction
\begin{equation}
\frac{n_n(T)}{n_e} = \left(\frac{T_s}{T_c}\right)^4.
\end{equation}

Taking the relation $\kappa_s \propto R_s$ for the quasiparticle induced losses and assuming $\omega_\mathrm{Res}, L_\mathrm{tot} \approx \mathrm{const.}$, which for this consideration is reasonable as their relative change is only $\sim 10^{-2}$, we get as cavity linewidth temperature dependence

\begin{equation}
\kappa(T) = \kappa_0 + \kappa_1\left(\frac{T_s}{T_c}\right)^4\cdot\left[1-\left(\frac{T_s}{T_c}\right)^4\right]^{-\frac{3}{2}}
\end{equation}
with a temperature independent contribution $\kappa_0$ and the scaling factor $\kappa_1$.
Figure~\ref{fig:Cavity_Zoom}b shows an approximation to the data using this expression with $\kappa_0 = 2\pi\cdot 850\,$kHz and $\kappa_1 = 2\pi\cdot 3.25\,$MHz ($T_\mathrm{off} = 1.09\,$K) as lines.

\subsection{Influence of the magnetic trapping fields}

Applying an external magnetic field can shift the cavity frequency as well as the cavity linewidth due to Meissner screening currents \cite{Healey08} and the presence of Abrikosov vortices \cite{Song09, Bothner12}.
In our experiment, we apply only small fields in the $100\,\mu$T range, but due to the fact that we also apply a field during the transition to the superconduting state, we will trap some vortices in the cavity leads \cite{Stan04}.
As the magnetic field distribution including vortices is very complicated for our device, we  describe the field-induced property shifts phenomenologically by slightly adjusting the kinetic inductance participation ratio $L_k/L_0$ and the parameter $\kappa_1$.
In Fig.~\ref{fig:Cavity_FullField}a, we plot the zero magnetic field data points and the analytic expressions (lines) as derived in the previous section and in \ref{fig:Cavity_FullField}b we show the experimental data obtained within the full magnetic trapping field configuration.
For comparison, we also plot the lines of \ref{fig:Cavity_FullField}a in \ref{fig:Cavity_FullField}b, but in grey, demonstrating that the magnetic fields indeed lead to a small resonance frequency downshift and a slight increase of the linewidth.
Both effects can be captured by using $L_k(B_\mathrm{trap})/L_0 = 0.02593$ and $\kappa_1(B) = 1.3\kappa_1(0)$.
The result is shown as black dashed lines in \ref{fig:Cavity_FullField}b and is in excellent agreement with the data.

\begin{figure}[h]
	\centering {\includegraphics[scale=0.75]{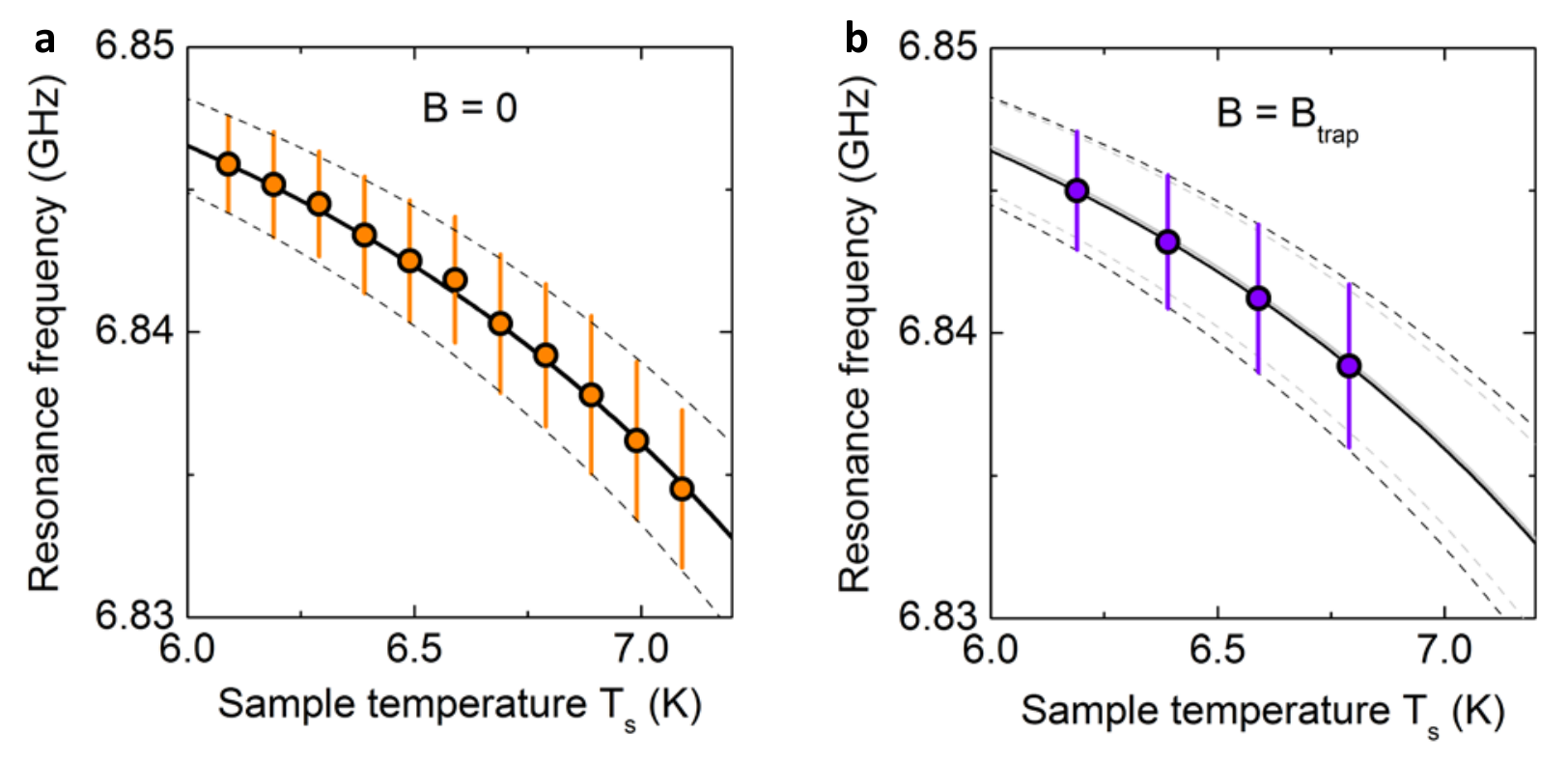}}
	\caption{\textsf{\textbf{Temperature and magnetic field dependence of the cavity parameters in the experimentally relevant range}. \textbf{a} Cavity resonance frequency and linewidth vs sample temperature in zero magnetic field. Circles show the resonance frequency values, bars on both sides of the points indicate the range $(\omega_\mathrm{Res}\pm\kappa)/2\pi$. Lines show corresponding analytical approximations as described in the main text. \textbf{b} Cavity resonance frequency and linewidth vs sample temperature with magnetic trapping fields applied. Circles show the resonance frequency values, bars on both sides of the points indicate the range $(\omega_\mathrm{cav}\pm\kappa)/2\pi$. Gray lines show the corresponding analytical approximations for $B=0$ as in a, and black lines indicate slightly modified expressions as described in the main text.}}
	\label{fig:Cavity_FullField}
\end{figure}

\section{Magnetic field simulations}

The magnetic field simulations in this work have been performed using the software package 3D-MLSI \cite{Khapaev03}.
For the calculations of the RF magnetic field, simplified versions of our real chip were used, as the full structure was too large to be computed to the full extent.
We do not expect the modifications (e.g. shortening the Z-shaped trapping wire to the trapping region), however, to have a significant impact onto the final results.

\subsection{Coupling per photon and atom}

The microwave current of the fundamental mode along the resonator is given by

\begin{equation}
I(l) = I_0\cos{\left(2\pi\frac{l}{\lambda_0}\right)}
\end{equation}
where $l$ is the coordinate along the resonator starting from the input port with $l = 0$, $\lambda_0 \approx 18.7\,$mm is the resonance wavelength and $I_0$ is the amplitude in the current antinodes.
To calculate the coupling rate $g$ between a single photon and a single atom in the cavity, we estimate the zero point fluctuations of the microwave current in the resonator and at the position of the atoms (current antinode) by

\begin{eqnarray}
\frac{1}{2}\hbar\omega_\mathrm{cav} & = & \int_0^{\lambda_0/2}L'I_\mathrm{zpf}^2\cos^2{\left(2\pi\frac{l}{\lambda_0}\right)}dl\\
& = & \frac{\lambda_0}{4}L'I_\mathrm{zpf}^2
\end{eqnarray}
where the inductance per unit length is $L' = 409\,$nH/m (kinetic inductance contributions are neglected here due to their smallness) and $I_\mathrm{zpf} = I_\mathrm{zpf0}/\sqrt{2}$ is the root mean square of the zero point fluctuation amplitude $I_\mathrm{zpf0}$.
With $\omega_\mathrm{cav} = 2\pi\cdot6.84\,$GHz and $\lambda_0 \approx 18.7\,$mm we get

\begin{equation}
I_\mathrm{zpf} = \sqrt{\frac{\hbar\omega_\mathrm{cav}}{\pi L'}} \approx 33.5\,\mathrm{nA}.
\end{equation}

To relate this to the coupling, we calculate the magnetic field $B_\mathrm{ph}$ related to this current at the position of the atoms by means of finite element simulations using the software package 3D-MLSI \cite{Khapaev03}.
Finally, we take into account the position of the atomic cloud along the resonator, which reduces the effective magnetic field to $\sim 0.95 B_\mathrm{ph}$.
Figure~\ref{fig:Supp_MW} shows the magnetic microwave field zero point fluctuations obtained from these simulations in a cross-section of the coplanar waveguide at the position of the atoms.
%
%

\begin{figure}[h]
	\centering {\includegraphics[scale=0.65]{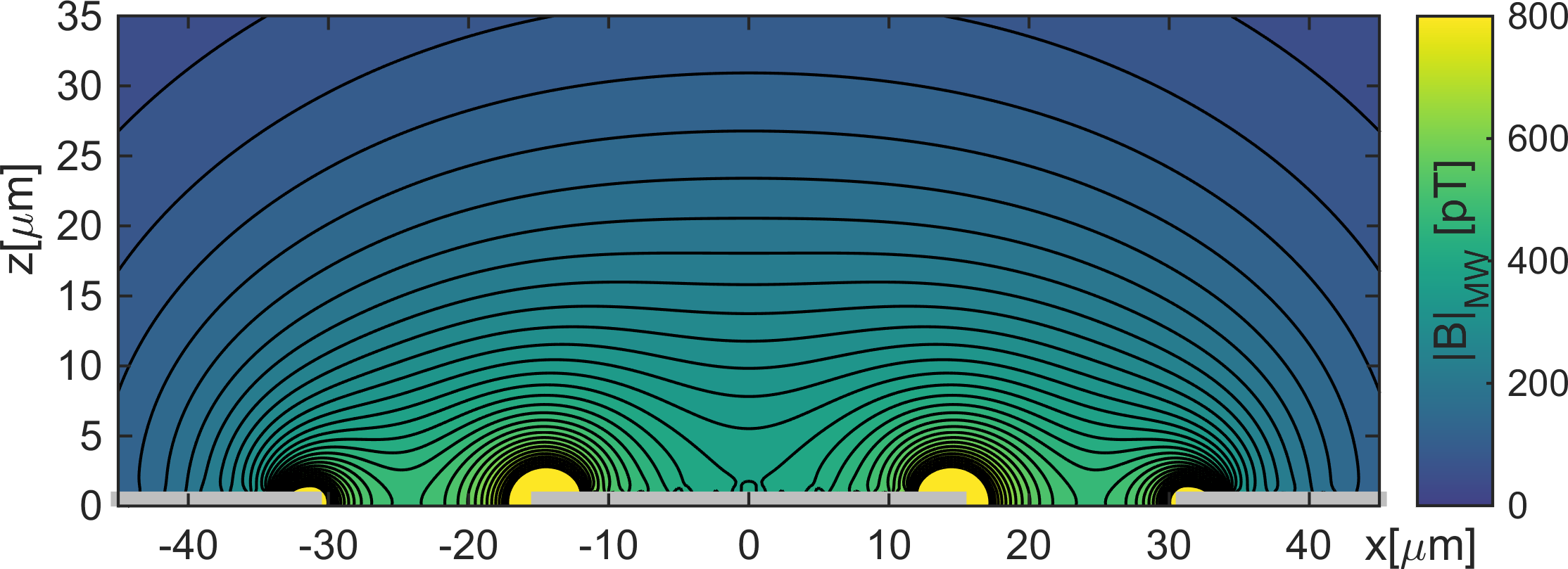}}
	\caption{\textsf{\textbf{Single-photon microwave magnetic field in the resonator}. The magnetic microwave field zero point fluctuation amplitude $|B|=|B_\mathrm{ph}|$ obtained by finite element simulations above the coplanar microwave structure. The coplanar waveguide structure is indicated by the grey bars at the bottom. The thickness of the CPW is not to scale.}}
	\label{fig:Supp_MW}
\end{figure}

From the magnetic microwave field, we calculate the single-atom coupling rate as

\begin{equation}
g = \frac{|B_\mathrm{ph}(x, y)|\cdot|\mu|}{\hbar}
\end{equation}
with the magnitude of the dipole transition matrix element $|\mu| =  0.25\mu_\mathrm{B}$.
The result is shown in Fig.~1f of the main paper.

\subsection{The radio-frequency magnetic field}

For the two-photon experiments and the corresponding simulations, we also need the magnetic field of the radio-frequency (RF) current, which is sent through the Z-shaped trapping wire.
Thus, we calculate the magnetic field for a current of $I_\mathrm{RF} = 1\,$mA on the trapping wire and show the result at the position of the atoms in Fig.~\ref{fig:Supp_RF}.

\begin{figure}[h]
	\centering {\includegraphics[scale=0.65]{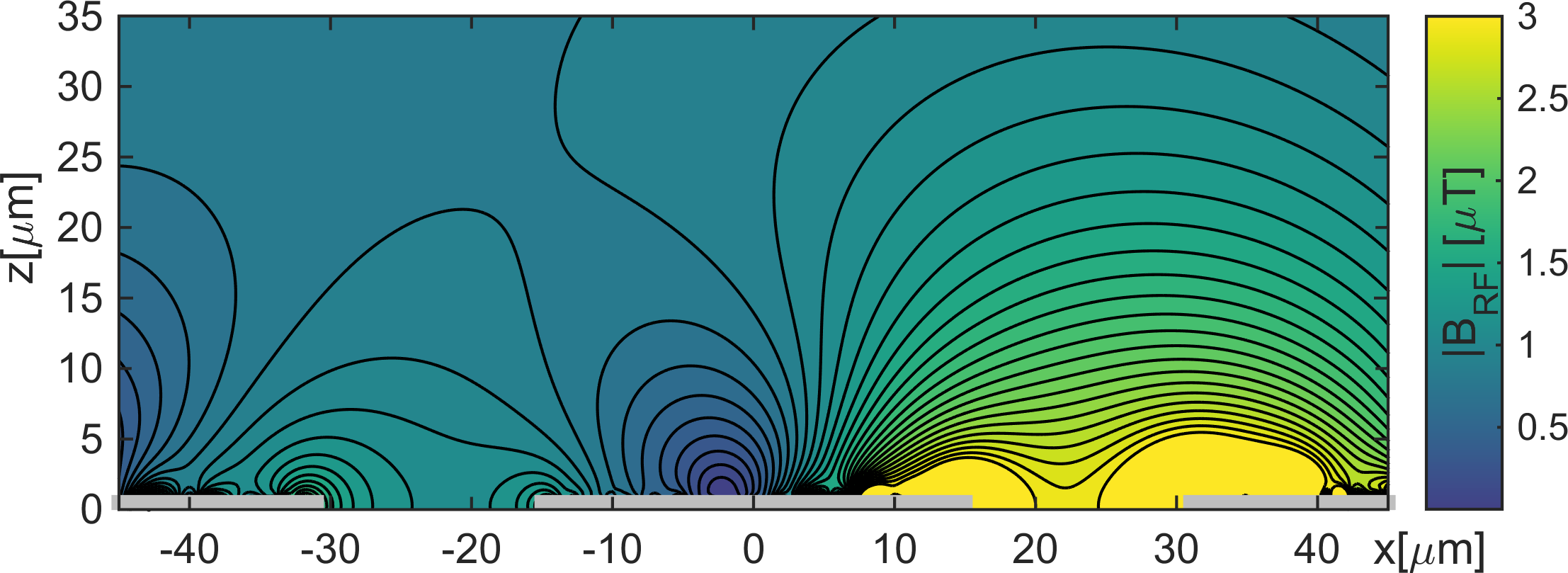}}
	\caption{\textsf{\textbf{Magnetic field of a current through the trapping wire in the resonator}. The plot shows the magnetic field generated in the microwave cavity, when a current of $I_\mathrm{RF} = 1\,$mA is flowing through the trapping wire. The grey bars at the bottom indicate the position of the resonator leads. The centre bar shows the centre conductor, the left bar corresponds to the $50\,\mu$m ground plane towards the trapping wire and the right bar corresponds to the large purely superconducting ground. The magnetic field is mainly guided through the right resonator gap due to the closed superconducting loop around the left gap (defined by centre conductor, ground and the coupling inductors).}}
	\label{fig:Supp_RF}
\end{figure}

\section{Simulated Rabi oscillations in the cavity}
\subsection{One photon Rabi oscillations}
Numerical simulations of the coherent Rabi oscillations of atomic ensembles in the cavity yield further insight into the observed dephasing rates. 
We assume a thermal ensemble of atoms with a temperature of $T_a =2000$\,nK trapped in a harmonic magnetic trap with $\omega_x = 2\pi\cdot 400$\,s$^{-1}$, $\omega_y = 2\pi\cdot 25$\,s$^{-1}$, $\omega_z = 2\pi\cdot 600$\,s$^{-1}$. 
The centre of the trap is assumed $20\,\mu$m from the chip surface, as depicted in Fig.\,1f in the main article.

For the one-photon Rabi oscillation, the Rabi frequency is much higher than the oscillation frequency of the atoms in the trapping potential, i.e. $\Omega_0 \gg \omega_z$.
We therefore can assume a static Gaussian density distribution of atoms in the trap, and use a total atom number of $1.2\times 10^5$ atoms for the simulations.
We use the numerically calculated field strength depicted in Fig.\,\ref{fig:Supp_MW}, multiplied by a constant numerical factor to match the observed Rabi oscillation frequency.
For each position $\vec{r}_i$, the probability to find atoms in the excited state is computed as
\begin{equation}
p_2(\vec{r}_i,t) = \frac{\Omega_0(\vec{r}_i)^2}{{\tilde\Omega(\vec{r}_i)^2}}\sin^2\left( \frac{{\tilde\Omega(\vec{r}_i)^2}}{2} t \right),
\end{equation}
where $\tilde\Omega(\vec{r}_i)^2 = \Omega_0(\vec{r}_i)^2 + \Delta(\vec{r}_i)$ is the generalized Rabi frequency, and $\Delta(\vec{r}_i)$ the magnetic-field dependent detuning of the microwave to the atomic transition.
The probability $p_2(\vec{r}_i,t)$ is multiplied with the local atomic density $n_\text{at}(\vec{r}_i)$ and summation over all atoms yields the total atom number in the excited state. 
The simulated results closely match the observed dephasing of the Rabi oscillations, as seen in Fig.\,3a in the main paper.

\subsection{Two-photon Rabi oscillations}
For the simulated two-photon Rabi oscillations, we assume a three level system of states $\left|1,-1\right\rangle$, $\left|2,0\right\rangle$, and $\left|2,1\right\rangle$. 

States $\left|1,-1\right\rangle$ and $\left|2,0\right\rangle$ are coupled by the cavity microwave field with the Rabi frequency $\Omega_\text{MW}$.
An additional radio frequency $\Omega_\text{RF}$ couples the state $\left|2,0\right\rangle$ to the state $\left|2,1\right\rangle$.
Both the microwave and the radio frequency field are detuned to the transition to the intermediate state $\left|2,0\right\rangle$ by the detuning $\pm\Delta$, c.f. Fig. 5b in the main article.
The inhomogeneity of the cavity field $\Omega_\text{MW}$ is the same as for the one-photon case above.
The spatial dependence of the radio-frequency field $\Omega_\text{RF}$ is simulated with the software package 3D-MLSI by applying a current in the Z-shaped wire and calculating the Meissner screening currents close to the resonator, c.f. \ref{fig:Supp_RF}. 
As the effective Rabi frequency is much lower as in the one-photon case, the assumption of static atoms no longer holds. 
The motion of atoms through the spatially inhomogeneous MW and RF field leads to a time dependence of the Rabi frequency seen by each atom. 

To account for this, we randomly initialize 5000 non-interacting particles in the state $\left|1,-1\right\rangle$ in the harmonic potential with a distribution corresponding to a temperature of 800\,nK. 
We then simulate the movement of the atoms through the potential and the evolution of the three states with a Runge-Kutta calculation of fourth order. 
Stability of the simulations was ensured by changing the time steps in the calculations.
The main source of the dephasing in the Rabi oscillations is the inhomogeneity of the MW field. 
This can be seen from simulations with colder and thus smaller clouds, as visible in Fig.\,\ref{fig:Supp_Coherence_T}.

\begin{figure}[!h]
	
	\centering
	\includegraphics[height = 4cm]{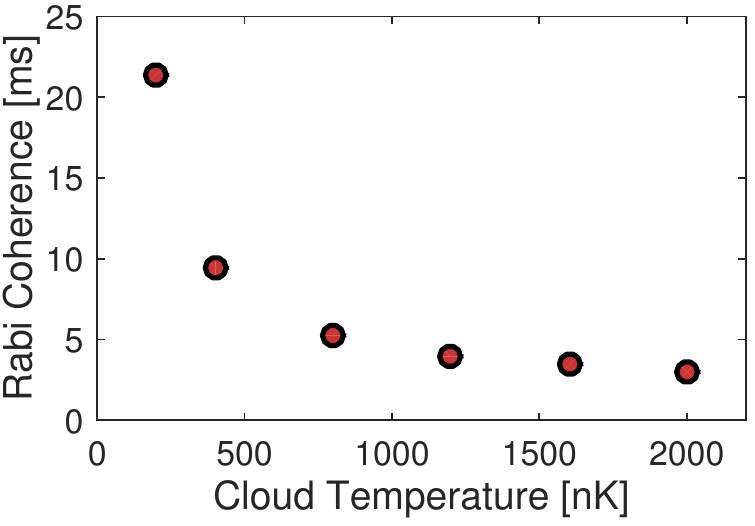} 
	\caption{\textsf{\textbf{Simulated coherence of the Rabi oscillations as a function of the cloud temperature.} The simulations assume trapping frequencies as in the experiment, i.e. $\omega_y = 2\pi\cdot 25$\,s$^{-1}$, $\omega_x = 2\pi\cdot 400$\,s$^{-1}$, $\omega_z = 2\pi\cdot 600$\,s$^{-1}$ . The high coherence for low temperatures shows that the inhomogeneity of the MW field is the primary source of the dephasing.}}
	\label{fig:Supp_Coherence_T}.
	
\end{figure}

\end{document}